\pdfoutput=1 

\documentclass{article}

\usepackage{arxiv}
\usepackage{physics}
\usepackage{qcircuit}
\usepackage[utf8]{inputenc} 
\usepackage[T1]{fontenc}    
\usepackage{hyperref}       
\usepackage{url}            
\usepackage{booktabs}       
\usepackage{amsfonts}       
\usepackage{nicefrac}       
\usepackage{microtype}      
\usepackage{bbold}
\usepackage{subcaption}
\usepackage{fullpage}
\usepackage{times}
\usepackage{enumitem}
\usepackage{fancyhdr,graphicx,amsmath,amssymb}
\usepackage{mathtools} 
\usepackage[ruled,vlined]{algorithm2e}
\usepackage{nicefrac}
\usepackage{floatrow}
\usepackage{xparse}
\usepackage{xcolor}

\newsavebox{\twosubbox}
\usepackage[titletoc,title]{appendix}

\usepackage{tabularx} 
\usepackage{multirow}

\newcommand{\appendixhead}%
{\textbf{\huge Appendices}
\vspace{0.25in}}

\newfloatcommand{capbtabbox}{table}[][\FBwidth]

\definecolor{maiblue}{rgb}{0, 0., 0.69}

\definecolor{Gray}{gray}{0.925}

\usepackage{graphicx}
\usepackage{tabularx} 
\usepackage{multirow}
\usepackage{booktabs}
\usepackage[mode=math]{siunitx}
\usepackage{colortbl}
\usepackage{booktabs}
\usepackage{hyperref}

\definecolor{upforestgreen}{rgb}{0.0, 0.27, 0.13}

\title{MAQA: A Quantum Framework for Supervised
Learning}

\author{
\href{https://orcid.org/0000-0002-1348-250X}{\includegraphics[scale=0.06]{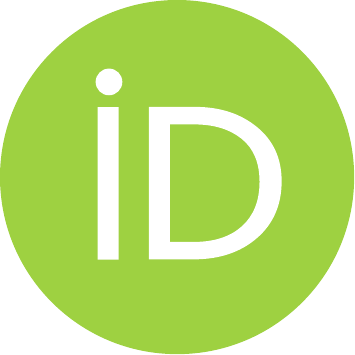}\hspace{1mm}Antonio Macaluso}\thanks{corresponding author} \\
	German Research Center for Artificial Intelligence \\(DFKI)\\
	Saarbruecken, Germany\\
	\texttt{antonio.macaluso@dfki.de} \\
	\And
        Matthias Klusch \\ 
	German Research Center for Artificial Intelligence \\(DFKI)\\
	Saarbruecken, Germany\\
	\texttt{matthias.klusch@dfki.de} \\
  \AND
      Stefano Lodi \\
    University of Bologna\\
    Bologna, Italy \\
    \texttt{stefano.lodi@unibo.it} \\
    \And
    Claudio Sartori \\
    University of Bologna\\
    Bologna, Italy \\
    \texttt{claudio.sartori@unibo.it} \\
}

\begin{document}

\maketitle

\begin{abstract}
Quantum Machine Learning has the potential to improve traditional machine learning methods and overcome some of the main limitations imposed by the classical computing paradigm. However, the practical advantages of using quantum resources to solve pattern recognition tasks are still to be demonstrated.

This work proposes a universal, efficient framework that can reproduce the output of a plethora of classical supervised machine learning algorithms exploiting quantum computation's advantages. The proposed framework is named  \textit{Multiple Aggregator Quantum Algorithm} (MAQA) due to its capability to combine multiple and diverse functions to solve typical supervised learning problems. 
In its general formulation, MAQA can be potentially adopted as the quantum counterpart of all those models falling into the scheme of aggregation of multiple functions, such as ensemble algorithms and neural networks. From a computational point of view, the proposed framework allows generating an exponentially large number of different transformations of the input at the cost of increasing the depth of the corresponding quantum circuit linearly. Thus, MAQA produces a model with substantial descriptive power to broaden the horizon of possible applications of quantum machine learning with a computational advantage over classical methods. As a second meaningful addition, we discuss the adoption of the proposed framework as hybrid quantum-classical and fault-tolerant quantum algorithm.

\end{abstract}

\keywords{Quantum Machine Learning \and Quantum Computing \and Machine Learning}

\section{Introduction}

Quantum computers are machines that leverage the properties of quantum mechanics to store and process information. 
Although a potential quantum advantage has already been shown in different fields, such as quantum chemistry \cite{peruzzo2014variational}, multi-agent systems \cite{venkatesh2022bilp,venkatesh2022gcs}, it is still unclear whether quantum computing can be  efficiently used for machine learning (ML) tasks.

The intersection between ML and quantum computing (QC) is known as Quantum Machine Learning (QML). 
There are two ways in which ML and QC can be combined: one approach is to run the learning process predominantly in a quantum computer so that the expensive subroutines can be executed efficiently. For this purpose, a rich collection of quantum algorithms for basic linear algebra subroutines have been proposed in literature \cite{biamonte2017quantum,nakahara2008quantum,dowson1978spectral}. 
Some popular examples of this approach are QSVM \cite{rebentrost2014quantum} and QSplines \cite{macaluso2020quantum}, which obtain an exponential speed-up with respect to their classical counterparts. However, the protocols within this category usually assume the availability of a fault-tolerant quantum computer.

Alternatively, variational quantum algorithms can be considered machine learning models that can be trained using hybrid quantum-classical optimization.
In this case, a quantum algorithm is used to make a call to a function that allows estimating the target variable of interest given the input data and a set of rotation parameters \cite{benedetti2019parameterized,schuld2018circuit}. This approach requires a parametrised quantum circuit and a classical optimisation procedure to find the optimal set of parameters for a sequence of quantum gates.
Although these techniques represent the most promising attempt to leverage near-term quantum technology, it is still unclear whether they can outperform classical algorithms.

Despite the remarkable success of ML in numerous real-world applications, the ever-increasing size of datasets and the high computational requirements of modern algorithms indicate that the current computational tools will no longer be sufficient in the future. 
In this work, we propose a novel and efficient quantum framework to reproduce a plethora of machine learning models using quantum computational advantages. The framework is called \textit{Multiple Aggregator Quantum Algorithm} (MAQA) due to its capability
to combine multiple and diverse functions to solve typical supervised learning tasks. Thanks to superposition, entanglement and interference, the MAQA framework can compute the weighted average of an exponentially large number of functions while increasing the depth of the correspondent quantum circuit linearly. This allows for building quantum models with incredible descriptive power that might be a credible alternative to classical methods in the future.

\section{Preliminaries}\label{cap3 sec: ML model as aggregator of functions}

The objective of a supervised model is to find a useful approximation to the function $f(x; \theta)$ that underlies the predictive relationship between the input $x$ and output $y$ for a fixed set of parameters $\theta$. Assuming for simplicity an additive error, the model of interest can be expressed as follows: 
\begin{align}\label{chapter3: eq ML standard formulation}
    y = f(x; \theta) +  \epsilon,
\end{align}
where $\epsilon$ is a random variable whose conditioned probability distribution given $x$ is centred in zero.
Although Eq. \eqref{chapter3: eq ML standard formulation} provides a general mathematical formulation for supervised learning, several methods do not estimate a single function but explicitly calculate multiple and diverse functions. These functions belong to the same family but differ in either a set of parameters or the training data. In all these cases, the final output results from the weighted average of the estimated functions:
%
\begin{align}\label{cap3 eq:ML model as aggregator}
    y = f(x;\theta) & = \sum_{h=1}^{H} \beta_h g(x;\theta_h), 
\end{align}
where $f(x; \theta)$ is the final output and $g(x; \cdot)$ describes the \textit{function component}.

The calculation of $g(x; \cdot)$ corresponds to a specific transformation of data $x$ based on $\theta_h$, whose contribution to the final output is weighted by $\beta_h$. The estimation of a collection of functions components allows producing an extremely flexible model, which is able to approximate the behaviour of complex patterns. Different choices for $\beta$, $g(x; \cdot)$ and $\theta_h$ determine different supervised models commonly adopted in real-world applications.

For instance, a single-layer neural network (or Single Layer Perceptron - SLP)  assumes as function component $g(x; \cdot)$ the activation function $\sigma_{\text{hidden}}$ applied to the linear combinations $L(x;\theta_h)$ of the input vector $x$. In fact, an SLP with $H$ hidden neurons is a two-stage model that takes as input training data $x$ and $H$ sets of linear coefficients and estimates the target variable as follows:
\begin{align}\label{eq:classical_SLP}
    f_{\text{SLP}}(x) = \sigma_{\text{output}}
    \left[\sum_{h=1}^{H} \beta_h \sigma_{\text{hidden}}\left(L(x;\Theta_h)\right)\right],
\end{align}
where $\sigma_{\text{output}}$ is the identity function when the task is the function approximation.
\footnote{When considering a neural network with multiple hidden layers, the only difference in Eq. \eqref{eq:classical_SLP} is that the function component $g(x; \cdot)$ is, in turn, a neural network.}

Another classical supervised learning approach that falls into the schema of function aggregation is ensemble learning. 
In practice, ensemble methods reduce to computing several  predictions $g_{1}(x), g_{2}(x), \dots , g_{H}(x)$ using $H$ different training sets, which are then averaged to obtain a single model:
\begin{align}
    f_{\text{ens}}(x) = \frac{1}{H} \sum_{h=1}^H \beta_h g_{h}(x).
\end{align}
In this case,
 the component functions $g(x;\cdot)$ are weak classification/regression models and the choice of the weights depends on the type of the ensemble in use (boosting, bagging, randomisation).

Other models that fit into the idea of multiple aggregations are Generalised Additive Models \cite{hastie1990generalized}, Support Vector Machines and Decision trees \cite{hastie01statisticallearning}.

\paragraph{Contribution.} In this work, we propose a novel efficient quantum framework to reproduce the idea of machine learning models as functions aggregators. The proposed architecture, named \textit{Multiple Aggregator Quantum Algorithm} (MAQA), can potentially reproduce some of the most important classical supervised learning algorithms introducing relevant computational advantages. In particular, MAQA propagates an input state to multiple quantum trajectories in superposition, and each trajectory describes a specific function $g(x; \cdot)$ that represents the component function of the final model. The entanglement between the two quantum registers involved (data and control) allows for efficient averaging of those transformations, and the final result can be accessed by measuring only a subset of qubits. The proposed approach has two main advantages: from a classical perspective, it introduces an exponential scaling in the number of aggregated functions while linearly increasing the time complexity of the correspondent quantum algorithm. From a quantum perspective, the framework opens the possibility of implementing a plethora of models not yet proposed in the literature.
Eventually, we discuss the adoption of MAQA to generalise some existing QML algorithms, considering both fault-tolerant settings and hybrid quantum-classical algorithms.

\section[Multiple Aggregator Quantum Algorithm (MAQA)]{Multiple Aggregator Quantum Algorithm (MAQA)}

In this section, we describe the MAQA framework that is able to reproduce the classical model expressed in Eq. \eqref{cap3 eq:ML model as aggregator}. The algorithm leverages the three main properties of quantum computing (superposition, entanglement and interference) to encode in a quantum state the sum of different input transformations accessible by measuring a single quantum register. 
The proposed algorithm can potentially reproduce all those models that refer to the idea of functions aggregation and provide attractive computational advantages with respect to the classical counterparts. 

The quantum algorithm adopts two quantum registers: data and control. The $data$ register encodes the model's input data, and the $control$ register is used to generate multiple trajectories in superposition, where each trajectory represents a different transformation of data. 

Starting from a $n$-qubit $data$ register and a $d$-qubit $control$ register the \textit{Multiple Aggregator Quantum Algorithm} (MAQA) 
involves four main steps: \textit{state preparation}, \textit{multiple trajectories in superposition}, \textit{transformation via interference} and \textit{measurement}. 

\paragraph{(Step 1) State Preparation \\}

\textit{State preparation} consists of encoding the input in the $data$ register and the initialisation of the $control$ register whose amplitudes depend on a set of parameters \textbf{$\beta=\{\beta^*_i\}_{i=1, \dots, 2^d}$}:
\begin{flalign}\label{eq: state_preparation}
    \ket{\Phi_0} & = (S_{\beta} \otimes S_{x}) \ket{0}_\text{control}^{\otimes d} \otimes \ket{0}^{\otimes n}_\text{data} \nonumber \\
    = & \frac{1}{\sqrt{2^d}}\sum_{h=1}^{2^d} \beta^*_h \ket{h} \otimes \ket{x}.
\end{flalign}
We refer to $S_{x}$ as a quantum routine to encode data into a quantum state, and $S_{\beta}$ as a routine that transforms a $d$-qubit register from an all-zero state to a quantum state which depends on a set of parameters 
$\beta$. 
Importantly, the computational cost of this step is not considered classically since any classical algorithm assumes the input $x$ to be directly accessible. 

\paragraph{(Step 2) Multiple Trajectories in Superposition \\}

The second step regards the generation of $2^d$ different transformations of the input data in superposition, each entangled with a possible state of the $control$ register. 
The single quantum state of the superposition encodes a specific transformation of the data and it depends on a set of parameters $\Theta_k$. To this end, a unitary $G(\theta_1, \dots, \theta_{2^d})$ that performs the following operation is assumed\footnote{Notice that the definition of $G(\theta_1, \dots, \theta_{2^d})$ unitary in terms of quantum gates depend on the specific algorithm in use (Sections \ref{MAQA as fault} and \ref{MAQA as hybrid}).}:
%
\begin{align}\label{eq:quantum aggregator}
     \ket{\Phi_1} & =  G\left(\theta_1, \dots, \theta_{2^d}\right) \ket{\Phi_0} 
     =  \frac{1}{\sqrt{2^d}} \left(\sum_{h=1}^{2^d}\beta^*_h\ket{h}\ket*{l(x;\Theta_h)}\right).
\end{align}
where the implementation of $G(\theta_1, \dots, \theta_{2^d})$ can be accomplished in only $d$ steps. 
Each step consists in the entanglement of the $i^{th}$ $(i=1, \dots d)$ \textit{control} qubit with two transformations $G\left(\theta_{i,1}\right)$ and $G\left(\theta_{i,2}\right)$ of $\ket{x}$ based on two sets of parameters,  $\theta_{i, 1}$ and $\theta_{i, 2}$.
Let us consider a unitary $G\left(\theta_{i,j}\right)$ that implements the transformation $l\left(x; \theta_{i,j}\right)$.
The most straightforward way to obtain the quantum state in Eq. \eqref{eq:quantum aggregator} is to apply $G\left(\theta_{i,j}\right)$ through controlled operations, using as \textit{control} state the two basis states of the current \textit{control} qubit. In particular, the generic $i^{th}$ step involves the following two transformations:

        First, the controlled-unitary $C^{(1)}G\left( \theta_{i,1}\right)$ is executed to entangle the transformation $G\left( \theta_{i,1}\right)\ket{x}$ with the excited state $\ket{1}$ of the $i^{th}$ \textit{control} qubit: 
            \begin{align}\label{eq:controlled-1}
                \hspace{-5em}
                \ket{\Phi_{i,1}}= & \left(C^{(1)}G\left( \theta_{i,1}\right)\right) \ket{c_i} \otimes \ket{x} 
                    = \left(C^{(1)}G\left( \theta_{i,1}\right)\right) \left(a_i\ket{0} 
                        + b_i\ket{1} \right) \otimes \ket{x} \nonumber \\
                    = & \left(a_i\ket{0}\ket{x} 
                        + b_i\ket{1} G\left( \theta_{i,1}\right)\ket{x} \right),
        \end{align}
        where $a_i$ and $b_i$ are the amplitudes of the $i^{th}$ \textit{control} qubit and $C^{(1)}G(\theta_{i,1})$ is a controlled operation that entangles the exited state of the \textit{control} qubit $\ket{c_i}$ to transform the $data$ register according to the unitary $G(\theta_{i,1})$. 
        %
 
        Then, a second controlled-unitary $C^{(0)}G\left( \theta_{i,2}\right)$ is executed. This time the \textit{control} state is the $\ket{0}$ basis state:
                \begin{align}\label{eq:controlled-0}
                \ket{\Phi_{i}}
                    = & \left( C^{(0)}G\left( \theta_{i,2}\right) \right)\ket{\Phi_{i,1}}  
                    = a_i\ket{0} G\left(\theta_{i,2}\right)\ket{x} 
                        + b_i\ket{1} G\left(\theta_{i,1}\right)\ket{x}.
                \end{align}
                

%
These two transformations are repeated for each qubit in the \textit{control} register and two different unitaries $G\left( \theta_{i,1}\right)$ and $G\left( \theta_{i,2}\right)$ are applied, at each iteration. 
After $d$ steps, the \textit{control} and \textit{data} registers are fully entangled and $2^d$ different quantum trajectories in superposition are generated. 
The output of this procedure can be expressed as follows:
\begin{align}
        \ket{\Phi_{d}}
        = & \frac{1}{\sqrt{2^d}} \sum_{h = 1}^{2^{d}} \beta^*_h \ket{h} G\left(\Theta_h\right)\ket{x} 
        =\frac{1}{\sqrt{2^d}} \sum_{h = 1}^{2^{d}} \beta^*_h \ket{h}\ket{l(x; \Theta_h)}
\end{align}
where $G\left(\Theta_h\right)$ results from the product of $d$ unitary matrices $G\left( \theta_{i,j}\right)$ and it represents a single quantum trajectory. Each trajectory differs from the others for, at least, one unitary $G\left( \theta_{i,j}\right)$\footnote{A detailed example with $d=3$ is described in Appendix \ref{sec:appendix}}.

When discussing a specific implementation of QML algorithms (Sections \ref{MAQA as fault} and \ref{MAQA as hybrid}), we will see that, from a computational point of view, the possibility to generate $2^d$ different transformations in only $d$ steps potentially leads to scaling exponentially the number of component functions with respect to classical methods, assuming an efficient implementation of the $C^{(j)}G(\theta_{i,j})$. 

\paragraph{(Step 3) Transformation via Interference \\}
Once we generated multiple transformations $l(x; \Theta_h)$ of the input in superposition, the third step consists of transforming the $data$ register through a generic quantum gate $F$ that works via interference:
\begin{align}\label{eq:transformation via interference}
    \ket{\Phi_{f}} 
                & = \left(\mathbb{1}^{\otimes d} \otimes F \right) \ket{\Phi_d} 
                = \left(\mathbb{1}^{\otimes d} \otimes F \right)\left[\frac{1}{\sqrt{2^d}}\sum_{h=1}^{2^d} \beta^*_h \ket{h} \ket{l(x; \Theta_h)}\right] \nonumber \\ 
                & = \frac{1}{\sqrt{2^d}}\sum_{h=1}^{2^d} \beta^*_h \ket{h} \ket*{g^*\left(x; \Theta_h\right)} 
                = \frac{1}{\sqrt{H}}\sum_{h=1}^{H} \beta^*_h \ket{h} \ket*{g^*_h} ,
\end{align}
where $H=2^d$.
In Eq. \eqref{eq:transformation via interference} the assumption is that the sequential application of $G(\Theta_h)$ and $F$ on the quantum state $\ket{x}$ is equivalent to calculate the function $g_h^*$ to an input $x$. 
At this point, different values of the function $g_h^*$ are entangled with different states of the $control$ register. 

It is important to notice that a single execution of $F$ allows the computation of the function $g_h^*$ for all the quantum trajectories in superposition. This is extremely useful when, during the computation, the same operations need to be applied to multiple inputs (e.g., when the activation function is applied to a huge number of neurons or in the case of ensemble learning, where the same classifier has to be executed to different sub-samples of the training set).


\paragraph{(Step 4) Measurement \\}
The last step consists of measuring the $data$ register, leaving untouched the $control$ register:
\begin{align}\label{eq: measurement}
    \left\langle M \right\rangle & =  \braket{\Phi_f|\mathbb{1}^{\otimes d} \otimes M}{\Phi_f} 
    =  
    \frac{1}{H}\sum_{h=1}^{H} \beta_h \braket{h}{h} \otimes \braket*{g^*_h|M}{g^*_h}  \nonumber \\
    & = \frac{1}{H} \sum_{h=1}^{H} \beta_h \braket*{g^*_h|M}{g^*_h} =
    \frac{1}{H} \sum_{h=1}^{H}\beta_h \left\langle M_h \right\rangle  \nonumber \\
    & = \frac{1}{H} \sum_{h=1}^{H} \beta_h  g\left(x; \Theta_h\right) = f_{\text{agg}},
\end{align}
 where $g\left(x; \Theta_h\right) =\braket*{g^*_h|M}{g^*_h}$ for a measurement operator $M$, $\beta_h = |\beta^*_h|^2$ with $\sum_h |\beta_h|^2 =1$ and $H=2^d$.

The expectation value $\left\langle M \right\rangle$ stores the weighted average of the $2^d$ functions $g\left(x; \Theta_h\right)$, which is accessible by measuring the $data$ register. 
While extracting the single contribution $g\left(x; \Theta_h\right)$ would require an exponential number of measurements (since those values are in the superposition of $2^d$ possible basis states), in a classical supervised learning scenario the measure of interest is the weighted average of all the functions which can be directly accessed by measuring the \textit{data} register and leaving intact the \textit{control} register. 

\begin{figure}[ht!]
    \centering
    \scalebox{.9}{\input{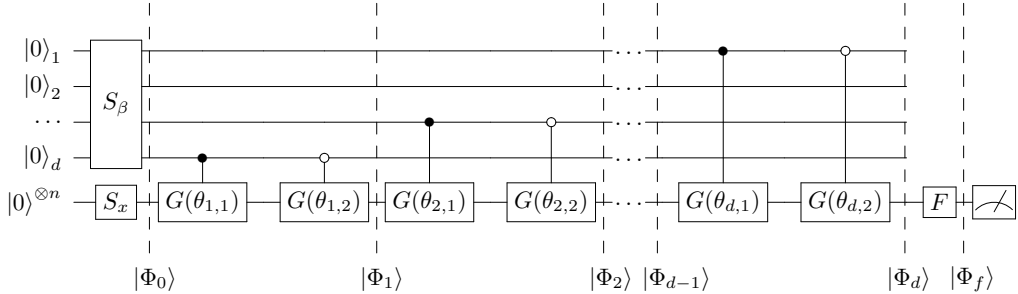}}
    \vspace{2em}
    \caption{Quantum Circuit for the \textit{Multiple Aggregator Quantum Algorithm} (MAQA).}
        \label{fig:circuit_MAQA}
\end{figure}

To summarise, the proposed architecture allows calculating the aggregation of multiple and diverse functions described in Eq. \eqref{cap3 eq:ML model as aggregator} using a quantum algorithm. In particular, it is possible to access the final result by measuring only the $data$ register while obtaining the weighted average of $2^d$ different transformations $g(x;\cdot)$ of the input data $x$, where $d$ is the size of the \textit{control} register.
Specifying properly $S_{\beta}$, $S_x$, $\{G\left(\theta_{i,1}\right), G\left(\theta_{i,2}\right)\}_{i=1, \dots, d}$ and $F$ allows potentially to reproduce the quantum version of all the ML algorithms discussed in Section \ref{cap3 sec: ML model as aggregator of functions}. 
Furthermore, the framework is very generic and can be adopted for hybrid and fault-tolerant quantum computation. The quantum circuit for implementing MAQA is depicted in Figure \ref{fig:circuit_MAQA}.

\section{Discussion}
As shown in the previous section, the MAQA allows obtaining a quantum state that reproduces the idea of ML models as aggregators of functions using the properties of quantum computing. From a classical ML perspective, relevant computational advantages are introduced.
Given $2^d$ component functions, any classical method that leverages the idea of functions aggregation scales linearly in $2^d$ since it is necessary to compute those functions explicitly to obtain the overall average. 
Furthermore, in the worst-case scenario, each component function has to process all available data; this implies a linear cost in the training set size multiplied by $2^d$. Using big-$\mathcal{O}$ notation, given a dataset $(x_i,y_i)$ for $i=1, \dots N$, where $x_i$ is a $p$-dimensional vector, and $y_i$ is the target variable of interest,  the overall time complexity of a model based on the aggregation of $2^d$ functions is: 
\begin{align*}
    \mathcal{O}(2^d N^{\alpha}p^{\beta}) \qquad \alpha, \beta \geq 1 .
\end{align*}
In contrast, MAQA generates a superposition of $2^d$ different transformations of the input in only $d$ steps since the single transformations are not computed directly, but they result from the combination of different unitaries $G(\theta_{i,j})$. 
Then, once the quantum state in Eq. \eqref{eq:transformation via interference} is generated, any operation  (unitary $F$) is propagated to all the quantum trajectories with a single execution. Using big-$\mathcal{O}$ notation, the time complexity of implementing the MAQA is:
\begin{align*}
    \mathcal{O}\left(d \times 2 C_G + C_F \right) ,
\end{align*}
where $C_G$ is the cost the controlled operation $C^{(j)}G(\theta)$ and $C_F$ is the cost of $F$. Note that the number of different functions grows exponentially with respect to the parameter $d$, which has a linear impact on the overall time complexity. This means that it is possible to generate an exponentially large number of different transformations of the input while obtaining their average efficiently, at the cost of increasing the depth of the corresponding quantum circuit linearly by a factor of $2C_G$. 

However, these advantages come with some compromises. First, the assumption about the nature of the operator $G(\theta_{i,j})$. In fact, MAQA assumes that the product of $G(\theta_{i,j})$ for $i=1, \dots , d$ 
produce a quantum gate $ G\left(\Theta_{k}\right)$:
\begin{align}\label{cap3: product G}
    G\left(\Theta_{k}\right) = \prod_{\substack{i=1, \dots, d\\j=1,2}} G\left(\theta_{i,j}\right) .
\end{align}
In practice, this means that multiple applications of the unitaries that depend on some set of parameters $\theta_{i,j}$ result in a single transformation of the same nature that depends on a derived set of parameters $\Theta_k$. Although any quantum circuit can be expressed as the product of different unitary matrices, the design of these gates in the context of supervised learning needs to be accomplished such that the final measurement provides the target variable of interest. 

Finally, when comparing classical and quantum algorithms, it is important to consider that quantum computation introduces a new complexity class, the \textit{Bounded-error Quantum Polynomial time}, representing the class of problems solvable in polynomial time by an innately probabilistic quantum Turing machine \cite{nielsen2002quantum}. Nevertheless, quantum algorithms need to be evaluated in terms of gate complexity. Thus, it is necessary that the exponential scaling introduced with respect to $d$ is preserved when considering a specific QML model.

\subsection{MAQA as Hybrid Quantum-Classical Algorithm}\label{MAQA as hybrid}
Recently the idea of aggregating two different unitary operators to reproduce the output of a two-neuron single-layer neural network via quantum circuit has been proposed (qSLP) \cite{macaluso2020variational,10.1007/978-3-031-25891-6_26}. 
Since multiple aggregations are the basis of both qSLP and MAQA, the latter can be seen as a natural extension of the former with an exponentially large number of neurons in the hidden layer. In fact, the entanglement between the \textit{control} and \textit{data} registers implies the number of linear combinations to be equal to the number of basis states of the \textit{control} register.
This, in turn, implies that the number of hidden neurons $H$ scales exponentially with the number of states of the \textit{control} register as a consequence of each hidden neuron being represented by a quantum trajectory.
This exponential scaling might enable the construction of a qSLP with an arbitrarily large number of hidden neurons as the amount of available qubits increases.
In other words, by adopting MAQA to generalise the qSLP, we can build a model with an incredible descriptive power capable of being a universal approximator.

From a computational point of view, given 
$H$ hidden neurons and $L$ training epochs, the training of a classical SLP scales (at least) linearly in $H$ and $L$ since the output of each hidden neuron has to be calculated explicitly to obtain the final output.  
Furthermore, if $H$ is too large (a necessary condition for an SLP to be a universal approximator \cite{macaluso2020variational,hornik1989multilayer}), the problem becomes NP-hard \cite{judd1990neural}. The adoption of MAQA to generalize the qSLP allows scaling linearly with respect to $log_2(H)=d$, thanks to the entanglement between the two quantum registers, which allows generating an exponentially large number of quantum trajectories in superposition.

However, the main challenge to tackle in the near future for qSLP-MAQA is still the design of a proper activation function -- in the sense of the Universal Approximation Theorem -- which is one of the significant issues for building a complete quantum neural network.
Yet, a recent proposal of QSplines \cite{macaluso2020quantum} opened the possibility of approximating non-linear activation functions via a quantum algorithm. Even so, QSplines use the HHL as a subroutine, a fault-tolerant quantum algorithm that cannot be adopted in hybrid computation on NISQ devices.

Nevertheless, recently it has been shown that quantum feature maps alongside functions aggregation is able to achieve universal approximation \cite{goto2020universal}. Thus, a possible future work consists of studying the qSLP-MAQA on top of the quantum feature map to enable it as a universal functions approximator without implementing a non-linear quantum activation function.
 
\subsection{MAQA as Fault-Tolerant Quantum Algorithm}\label{MAQA as fault}

Recently, a quantum algorithm that implements the idea of ensemble methods has been proposed \cite{macaluso2020ICTCS} and further developed  \cite{macaluso2020ensemble}. Looking at the specific quantum circuit in use, it is possible to observe that quantum ensembles can be considered as a particular instance of MAQA, where the controlled-rotation in Eq. \eqref{eq:controlled-0}, \eqref{eq:controlled-1}  are implemented using only the basis state $\ket{1}$ as \textit{control}  state which is transformed through \textit{Pauli}-X gate  at each iteration. Furthermore, while MAQA allows flexible quantum trajectories in terms of parametrised quantum gates $S_{\beta}$ and $\{G\left(\theta_{i,1}\right), G\left(\theta_{i,2}\right)\}_{i=1, \dots, d}$, in the case of quantum ensemble \cite{macaluso2020ensemble} the weights are pre-fixed (uniform superposition of the \textit{control} register) and the transformations of the input data are represented by CSWAP operations.
Thus, MAQA potentially extends the proposed quantum ensemble specifically defined for bagging strategy to other ensembles such as boosting and randomisation, where the parameters of the single base model and the correspondent weights are not pre-fixed.
Still, the main drawback of the quantum ensemble remains the underlying assumption to encode the complete training and test set into two different quantum registers and use a large number of trajectories in superposition to compute different subsamples of the training set. This would require an incredibly high number of qubits in a fault-tolerant quantum computer. 

In this respect, the main challenge to tackle to make the ensemble effective (using MAQA) in the near future is the design of a quantum classifier based on interference that guarantees a more efficient data encoding strategy (e.g. amplitude encoding) and can process larger datasets.

\section{Conclusions}
The practical advantages of using quantum resources to solve machine learning tasks are still to be demonstrated. However, the ground provided by quantum mechanics is highly appealing since a low number of qubits allows accessing an exponentially large Hilbert space. 

In this work, we tried to take a further step towards the study of how machine learning can benefit from quantum computation.
The proposed quantum framework, the \textit{Multiple Aggregator Quantum Algorithm} (MAQA), is capable of reproducing  some of the most important classical machine learning algorithms using quantum computing resources.
MAQA can potentially improve, in terms of time complexity, all those models that require explicitly computing multiple and diverse functions to produce a final strong model. 
In particular, the cost aggregating $H$ different functions in classical machine learning requires a computational cost linear in $H$. Instead, the proposed quantum architecture allows scaling exponentially in $H$, requiring only $log_2(H)$ steps under the assumption that the cost in terms of circuit complexity is unitary for each step.
The advantage comes directly from using superposition and entanglement as resources for generating different transformations of the input. 
Furthermore, quantum interference allows propagating the use of a specific unitary (gate $F$) to all the quantum trajectories in superposition. Hence, the application of $F$ impacts additively the overall time complexity, and the same operation would require a multiplicative cost in classical computation. 

In addition, we discussed how the proposed approach could be adopted as a fault-tolerant (quantum ensemble) and hybrid quantum-classical (quantum Single Layer Perceptron) algorithm, though different technical aspects need to be further investigated for both cases.

We are still in an early stage of QML, and its contribution to solving real-world in the context of machine learning is yet to be understood. However, many research findings, including this work, suggest that the potential of quantum computing is huge, and machine learning will likely benefit from it in the future.

\section*{Acknowledgment}
This work has been partially funded by the German Ministry for Education and Research (BMB+F) in the project QAI2-QAICO under grant 13N15586.

This version of the article has been accepted for publication, after peer review
but is not the Version of Record and does not reflect post-acceptance improvements, or any
corrections. The Version of Record is available online at: \href{http://dx.doi.org/10.1007/s11128-023-03901-w}{http://dx.doi.org/10.1007/s11128-023-03901-w}.

\section*{Additional Information}
This article is based on Chapter four of the author's PhD thesis \cite{macaluso2021novel}.

\newpage



\bibliographystyle{elsarticle-num} 

\bibliography{references}

\newpage

\section*{Appendix}\label{sec:appendix}
We consider the MAQA architecture when $d=3$. To make the notation simpler, we indicate the parametrised quantum gate $G(\theta_{i,j})$ as $G_{i,j}$. 

Without loss of generality, we can express the quantum gate $S_{\beta}$ as the tensor product of $d=3$ unitary gates $B_i$. Then, the state preparation step can be expressed as follows:
\begin{align}
\hspace{-3em}
    \ket{\Phi_0}& = (S_{\beta} \otimes S_{x}) \ket{0, 0 , 0}_{\text{control}} \otimes \ket{0, \dots, 0}_{\text{data}} \nonumber \\
                & = \left( \overset{d}{\underset{i=1}{\otimes}} B_i \otimes S_{x} \right)  \ket{0}^{\otimes 3} \otimes \ket{0}^{\otimes n} \nonumber \\
                & = \overset{3}{\underset{i=1}{\otimes}} \left(a_{i} \ket{0} + b_{i} \ket{1}\right) \otimes \ket{x} = \overset{3}{\underset{i=1}{\otimes}} \ket{c_{i}}\otimes \ket{x},
\end{align}
where $\ket{c_{i}}$  is the $i^{th}$ \textit{control} qubit and $a_i$ and $b_i$ are the parameters that determine its amplitudes:%
\begin{align}
    \ket{c_i} = a_{i} \ket{0} + b_{i} \ket{1}.
\end{align}
%

Once the two registers are initialised, each qubit in the $control$ register is entangled with two different random transformations of the $data$ register. 
Thus, 
the first step after state preparation is the following:

\paragraph{\bf{Step 1} ($i=1$) \\}

First, the controlled-unitary $C^{(1)}G_{1,1}$ is executed to entangle the transformation $G_{1,1}\ket{x}$ with the excited state of $\ket{c_3}$: 
    \begin{align}
    \hspace{-3em}
    \ket{\Phi_{0,1}} & = \left[ \mathbb{1}^{\otimes 2} \otimes C^{(1)}G_{1,1}\right] \ket{\Phi_0} \nonumber \\ 
                     & = \left[ \mathbb{1}^{\otimes 2} \otimes C^{(1)}G_{1,1}\right] \left( a_{3} \ket{0} + b_{3} \ket{1} \right) \otimes \ket{x} \nonumber \\
                    & = \ket{c_1} \otimes \ket{c_2} \otimes \left(a_{3}\ket{0}\ket{x} + b_{3}\ket{1} G_{1,1}\ket{x} \right).
    \end{align}

Then, a second controlled-unitary $C^{(0)}G_{1,2}$ is executed:
            \begin{align}
    \hspace{-3em}
            \ket{\Phi_{1}} & = \left[ \mathbb{1}^{\otimes 2} \otimes C^{(1)}G_{1,2}\right] \ket{\Phi_{1,1}} \nonumber \\
                          & = \overset{2}{\underset{i=1}{\otimes}} \ket{c_i}\otimes \left(a_3\ket{0}G_{1,2}\ket{x} + b_{3}\ket{0} G_{1,1}\ket{x} \right).
\end{align}

At this point, two different transformations, $G_{1,1}$ and $G_{1,2}$ of the initial state $\ket{x}$ are generated in superposition and are entangled with the two basis states of the \textit{control} qubit $\ket{c_3}$.

\paragraph{\bf{Step 2} ($i=2$) \\}
The same operations are applied using $\ket{c_2}$ as control qubit and different matrices, $G_{2,1}$ and $G_{2,2}$.

First, the controlled-unitary $C^{(1)}G_{2,1}$ is applied to entangle a transformation of $\ket{x}$ with the excited state of $\ket{c_2}$: 
            \begin{align}
            \ket{\Phi_{2,1}} & = 
            \left(\mathbb{1} \otimes C^{(1)} \otimes \mathbb{1} \otimes G_{2,1}\right) \ket{\Phi_{1}} \nonumber \\
            &  = \ket{c_1} \otimes \frac{1}{\sqrt{4}}\left[ a_{2}\ket{0}
            \left(a_3\ket{0}G_{1,2}\ket{x} + b_{3}\ket{0} G_{1,1}\ket{x} \right) 
            \right] + \nonumber \\ 
            & \hspace{4.9em} + b_2\ket{1} \left(a_3\ket{0}G_{2,1}G_{1,2}\ket{x} + b_3\ket{1} G_{2,1}G_{1,1}\ket{x} \right) ,
            \end{align}
    where the position of the gate $C^{(1)}$ indicates the \textit{control} qubit used for $G_{2,1}$.
Then, a second controlled-unitary $C^{(0)}G_{2,2}$ is executed:
\begin{align}\label{eq: appendix final}
        \ket{\Phi_{2}} & = \left( \mathbb{1} \otimes C^{(1)} \otimes \mathbb{1} \otimes G_{2,2}\right) \ket{\Phi_{2,1}} \nonumber \\ &
        = \ket{c_1} \otimes \frac{1}{\sqrt{4}}\Big[ a_2\ket{0} \big(a_3\ket{0}G_{2,2}G_{1,2}\ket{x} 
            + b_3\ket{1} G_{2,2}G_{1,1}\ket{x}\big) 
            \nonumber \\ & 
            \hspace{5em} + b_2\ket{0} \big(a_3\ket{0}G_{2,1}G_{1,1}\ket{x} + 
            b_3\ket{1} G_{2,1}G_{1,2}\ket{x} \big) \Big].
            \end{align}
Notice that the entanglement performed in \textbf{Step 2.1} influences the entanglement in \textbf{Step 2.2}, and  each trajectory describes a different transformation of $\ket{x}$. Eq. \eqref{eq: appendix final} can be rewritten expressing the four basis states of the control register using natural numbers:
\begin{align}
\ket{\Phi_{2}} = \ket{c_1} \otimes &
\frac{1}{\sqrt{4}}
\Big[
a_2a_3 \ket{00} G_{2,2}G_{1,2}\ket{x}    
+ 
a_2b_3\ket{01} G_{2,2}G_{1,1}\ket{x} 
\nonumber \\ & 
\hspace{1em} + b_2a_3\ket{10} G_{2,1}G_{1,2}\ket{x}
+
b_2b_3\ket{11} G_{2,1}G_{1,1}\ket{x} 
         \Big]  \nonumber \\
 & \hspace{-3.8em} =  \ket{c_1} \otimes \frac{1}{\sqrt{4}} \sum_{h=1}^{4} \beta^*_h \ket{h} G(\Theta_{h})\ket{x} ,
\end{align}
where $G(\Theta_{h})$ is the product of $d=2$ unitaries $G_{i,j}$, the coefficients $\beta^*_h$ result from the product of two coefficients $a_i$ and $b_i$. 
Thus, using $2$ control qubits $4$ different quantum trajectories are generated that correspond to $4$ different transformations of data $\ket{x}$.

\paragraph{\bf{Step 3} ($i=3$) \\}

Extending the same procedure when $d=3$, the result is the following:
\begin{align}
\hspace{-2em}
    \ket{\Phi_{3}}
    =  \frac{1}{\sqrt{8}}\Big[ 
    & 
    \beta^*_1 \ket{000} G_{3,2}G_{2,2}G_{1,2}\ket{x}
    +
    \beta^*_2 \ket{001} G_{3,2}G_{2,2}G_{1,1}\ket{x} 
    \nonumber \\ + & 
    \beta^*_3\ket{010} G_{3,2}G_{2,1}G_{1,2}\ket{x}
    +
    \beta^*_4\ket{011}G_{3,2}G_{2,1}G_{1,1}\ket{x}    
    \nonumber \\ + & 
    \beta^*_5\ket{100} G_{3,1}G_{2,2}G_{1,2}\ket{x}
    +
    \beta^*_6\ket{101} G_{3,1}G_{2,2}G_{1,1}\ket{x}    
    \nonumber \\ + & 
    \beta^*_7\ket{110} G_{3,1}G_{2,1}G_{1,2}\ket{x}
    +
    \beta^*_8\ket{111}G_{3,1}G_{2,1}G_{1,1}\ket{x}
    \Big] \nonumber \\
 & \hspace{-2.8em} = \frac{1}{\sqrt{8}} \sum_{h=1}^{8} \beta^*_h \ket{h} G(\Theta_{h})\ket{x} ,
\end{align}
where each $G(\Theta_{h})$ is the product of $3$ unitaries $G_{i,j}$ for $i=1,2,3$ and $j=1,2$.  

Repeating this procedure $d$ times with different control qubits the result is the following quantum state:
\begin{align}\label{cap3 eq:MAQA result}
        \ket{\Phi_{d}}
        & = \frac{1}{\sqrt{2^d}} \sum_{h = 1}^{2^{d}} \beta^*_h \ket{h} G(\Theta_{h})\ket{x} 
        = \frac{1}{\sqrt{2^d}} \sum_{h = 1}^{2^{d}} \beta^*_h \ket{h}\ket{l(x; \Theta_{h})} ,
\end{align}
where each $G(\Theta_{h})$ is the product of $d=3$ unitaries $G_{i,j}$ for $i=1, \cdots, d$ and $j=1,2$.

Finally, gate $F$ is applied, as shown in Eq. \eqref{eq:transformation via interference} and the measurement of the data register is performed.




\end{document}